\documentclass[12pt]{iopart}
\usepackage{float}
\usepackage{graphicx}
\usepackage[UKenglish]{babel}

\begin{document}
	\title{Thermal state quantum key distribution}
	\author{A Walton$^{1,3}$, A Ghesqui\`ere$^1$, G. Brumpton$^1$, D Jennings$^{1,2}$ and B Varcoe$^1$}
	\date{\today}
	\address{$^1$ School of Physics and Astronomy, University of Leeds, Leeds, LS2 9JT, United Kingdom}
	\address{$^2$ Department of Physics, Imperial College London, London, SW7 2AZ, United Kingdom}
	\address{$^3$ Email: pyaw@leeds.ac.uk}
	\begin{abstract}
		We analyse a central broadcast continuous variable quantum key distribution protocol in which a beam produced by a thermal source is used to create a secret key between two parties, Alice and Bob. A beam splitter divides the initial beam into a pair of output beams, which are sent to Alice and Bob, with Eve intercepting Bob's beam. We investigate the protocol in detail, calculating mutual informations through a pair of analytic methods and comparing the results to the outputs of a Monte Carlo simulation of the protocol. In a lossless system, we find that a lower bound on the key rate remains positive in the protocol under a beam splitter attack, provided Bob receives a nonzero proportion of the beam initially sent to him. This suggests that the thermal state protocol could be used experimentally to produce secure keys.
	\end{abstract}
	\maketitle
\section{Introduction}
In quantum key distribution (QKD), two parties, Alice and Bob, want to communicate in a secure fashion despite the presence of Eve, who is eavesdropping on their communication channel. They do this through establishing a cryptographic key that is known only to them and no one else \cite{Experimental_QKD,BB84}. However, Alice cannot simply send Bob a key over their communication channel, as Eve will also learn the key by eavesdropping. Therefore, protocols are needed which can distribute an identical key to Alice and Bob over an insecure channel, without Eve discovering it.

Currently, protocols exist that can accomplish this, though many methods of classical encryption base their security on the fact that certain mathematical operations, such as factorising large semiprime numbers, are very difficult to perform using current technology \cite{RSA}. However there is no reason to assume that solving these problems within a reasonable timeframe will continue to be difficult in the future as computing power increases and new algorithms are created.

In quantum protocols on the other hand, we make the assumption that Eve has access to arbitrarily large amounts of computing power while still being able to establish secure communication between Alice and Bob. This is done by basing security on restrictions imposed by the laws of quantum mechanics \cite{BB84}, such as the inability to measure a quantum state without affecting the system. This cannot be overcome through any amount of computing power.

Currently, most QKD protocols use coherent light, produced by lasers, as a method of generating secure keys. An example of this is the Gaussian Modulated Coherent State (GMCS) protocol \cite{GMCS_QKD,ExperimentalPassive,Passive}, where the key is encoded in randomly chosen quadratures of a beam described by randomly distributed coherent states. However, recently more analysis has been done concerning the use of thermal states in QKD \cite{Thermal_1,Thermal_2,Thermal_3}. These involve splitting a beam emitted by a thermal source at a beam splitter and sending the outputs to Alice and Bob respectively. Previous work concerning thermal states showed that they exhibit Hanbury Brown and Twiss correlations \cite{HBT} when split at a beam splitter, and quantum discord, a requirement for quantum key distribution \cite{Discord}.

One of the main factors limiting thermal methods is that noise and thermalisation of states are seen as detrimental for QKD protocols \cite{Noisy_QKD}, however work in this area is valuable due to the widespread use of microwaves in wireless modern communication, such as in WiFi and Bluetooth, in which thermal state QKD could be applied. Coherent state QKD is not suitable for these applications as the devices involved do not broadcast such states.

Here, we analyse a central broadcast protocol using a thermal input, with Eve intercepting the beam sent to Bob in order to eavesdrop. Monte Carlo simulations of the protocol are performed to produce sample bit strings, setting up for future experimental work using microwave sources. The paper begins with a brief overview of thermal states in Section \ref{sec:Thermal-states}, followed by Section \ref{sec:Protocol}, which describes the setup that will be simulated, while Sections \ref{sec:Information-Measurements}-\ref{sec:Covariance-Matrices} describe the measurements and workings.

\section{Thermal states} \label{sec:Thermal-states}

When written in the Fock basis, with \(\hat{a}^{\dagger}\) denoting the creation operator and \(|n\rangle=\frac{\left(\hat{a}^{\dagger}\right)^{n}}{\sqrt{n!}}|0\rangle\) describing an n-photon state, thermal states are given in the form \(\rho_{\mbox{\tiny Th}}=\sum_{n=0}^{\infty}p_{n}|n\rangle\langle n|\). Here, \(p_{n}=\frac{\exp\left(-n\beta\hbar\omega\right)}{1-\exp\left(-\beta\hbar\omega\right)}\) describes a thermal distribution where \(\beta=(k_{B}T)^{-1}\) is the thermodynamic beta. When a beam from a thermal source is input into a beam splitter, correlations are observed in intensity measurements performed on the output beams \cite{Thermal_1,HBT,Photon_Bunching} which are not present when a coherent source is used. 

These correlations exist due to the bunched nature of photons in thermal light. When detecting light from a thermal source, photons are not detected in random intervals, but are instead detected in clusters \cite{Photon_Bunching}. High variance in the intensity of thermal light, which is not present with a coherent source, is the result of this bunching.

We aim to take advantage of the correlations produced by this phenomenon to devise a QKD protocol which produces correlated bit strings between Alice and Bob using microwave sources. These bit strings can then be used to create a secure key to allow private communication in the presence of an eavesdropper.

The use of thermal states differentiates this protocol from similar versions involving modulated coherent states. Using a thermal source lets us carry out the protocol with common microwave-based wireless communication equipment instead of relying on fibre.

Additionally, the output of a thermal source and a Gaussian modulated coherent source are statistically equivalent. This allows the application of security proofs for GMCS protocols to thermal protocols. An important distinction to note is that coherent states are superpositions of Fock states, whereas thermal states are a mixture. This allows a Monte Carlo simulation to be used as an appropriate method to model the protocol, through random sampling of Fock states. Here, we will compare the outputs of such a simulation to mutual information values predicted through two separate analytic methods.

\section{The QKD Protocol} \label{sec:Protocol}

\begin{figure}[H]
	\centering
		\includegraphics[width=0.6\textwidth]{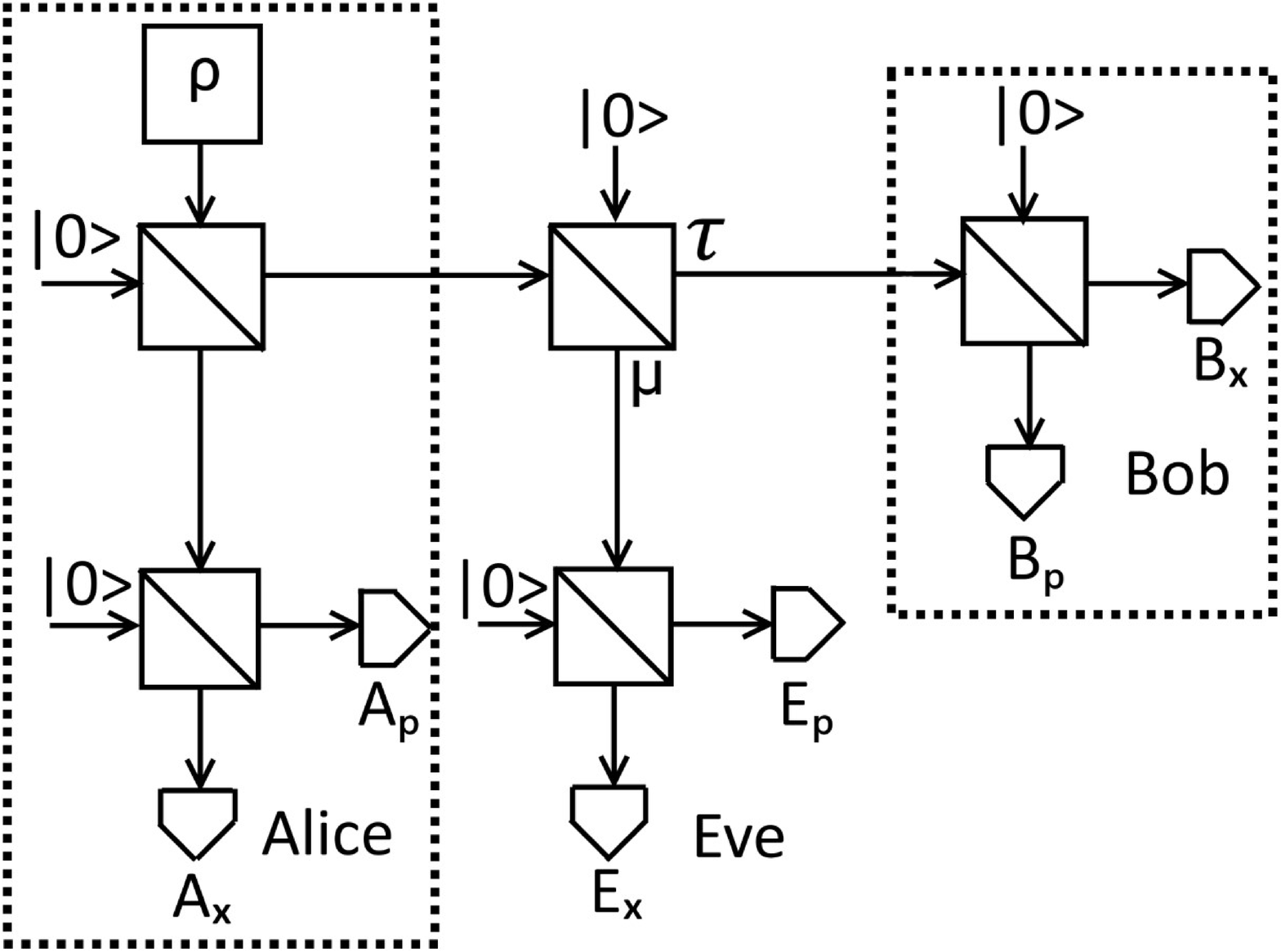}
	\caption{\textbf{Protocol Schematic.} A beam produced by a thermal source provides the initial state $\rho$. A series of beam splitters are used to direct the beam to Alice and Bob, with Eve performing a beam splitter attack on the channel leading to Bob. Eve's beam splitter has unknown transmittance and reflectance, $\tau$ and $\mu$, while each other beam splitter is 50:50.\label{fig:Protocol}}
\end{figure}

In the QKD protocol to be analysed, as shown in Figure \ref{fig:Protocol}, we use a central broadcast system in which light from a thermal source is incident on a 50:50 beam splitter. The output beams from this splitter are sent to Alice and Bob. An eavesdropper, Eve, uses a beam splitter attack, intercepting the beam sent to Bob using their own beam splitter of unknown transmittance. The part of the beam transmitted by Eve's beam splitter continues to Bob.

Alice is considered to be in control of the initial source, the first beam splitter, and the channels between the source and her measurement apparatus, while Bob is in control of their beam splitter and its output channels. The channel between the initial beam splitter and Bob is not under Alice or Bob's control, giving a point in the protocol where Eve may interfere with the system.

\begin{figure}[H]
	\centering
	\includegraphics[width=0.7\textwidth]{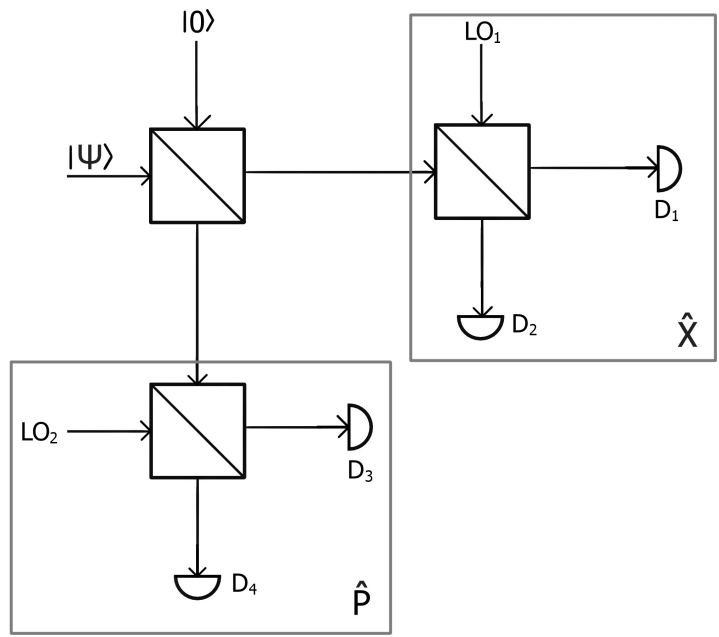}
	\caption{\textbf{Heterodyne Detection.} Heterodyne; or double homodyne, detection. As initially shown in Figure \ref{fig:Protocol}, the measurer splits the incoming signal at a 50:50 beam splitter, and combines the outputs with local oscillators. The X and P quadratures can then be measured separately using two pairs of detectors. \label{fig:DoubleHomodyne}}
\end{figure}

When each person receives their beam, they use a 50:50 beam splitter to divide the incoming signal into two outputs. Double homodyne detection is employed in order to measure the X quadrature of one beam, and the P quadrature of the second beam as shown in Figure \ref{fig:DoubleHomodyne}. Each person cannot simply measure the X and P quadratures of the single beam they receive as the quadrature operators do not commute. This replaces the common method of measurement in QKD, in which the variable to measure \cite{Ralph_1999} (or the measurement basis \cite{BB84}) is randomly switched in order to ensure security. This method of performing measurements in QKD without random basis switching has been previously used with success for continuous variable QKD protocols \cite{Basis}.

Repeated measurements yield an array of X and P quadrature measurements for each person. For each pair of quadrature measurement outcomes \(\left\{ x_{i},p_{i}\right\}\), Alice, Bob and Eve each calculate \(z_{i}=\sqrt{x_{i}^{2}+p_{i}^{2}}\), producing a distribution of $z$ measurements for each person. This is converted into a bit string by having Alice, Bob, and Eve each find the median value of their distribution, and recording a 0 or a 1 for each $z$ value depending on if it above or below the median. Due to the correlations in the outputs of the beam splitters with a thermal input, this produces a string of correlated bits for each person. At this point, if the protocol has been successfully executed, a key may be distilled from the bit strings, allowing Alice and Bob to communicate securely. Comparing Alice and Bob's results for a subset of measurements allows them to calculate correlation coefficients, to verify that a thermal source was used and correlated bits have been transmitted.

We performed a Monte Carlo simulation of this protocol in Python with QuTiP \cite{QuTiP2,QuTiP1}.
The initial beam is created by randomly sampling Fock state values from the thermal state distribution, with the beam splitters randomly splitting an input beam into a pair of outputs. With a Fock state input, the possible output Fock states of one arm of a beam splitter is described by a binomial distribution. One of these possible outputs is selected at random. This describes a portion of the incident photons being transmitted through the beam splitter, with the remaining portion being reflected. Once all the beam splitters are applied, each person receives a string of randomly distributed Fock state measurements. Due to thermal states being a statistical mixture of Fock states, this is an appropriate method of modelling the system.

\begin{figure}[H]
	\centering
	\includegraphics[width=0.8\textwidth]{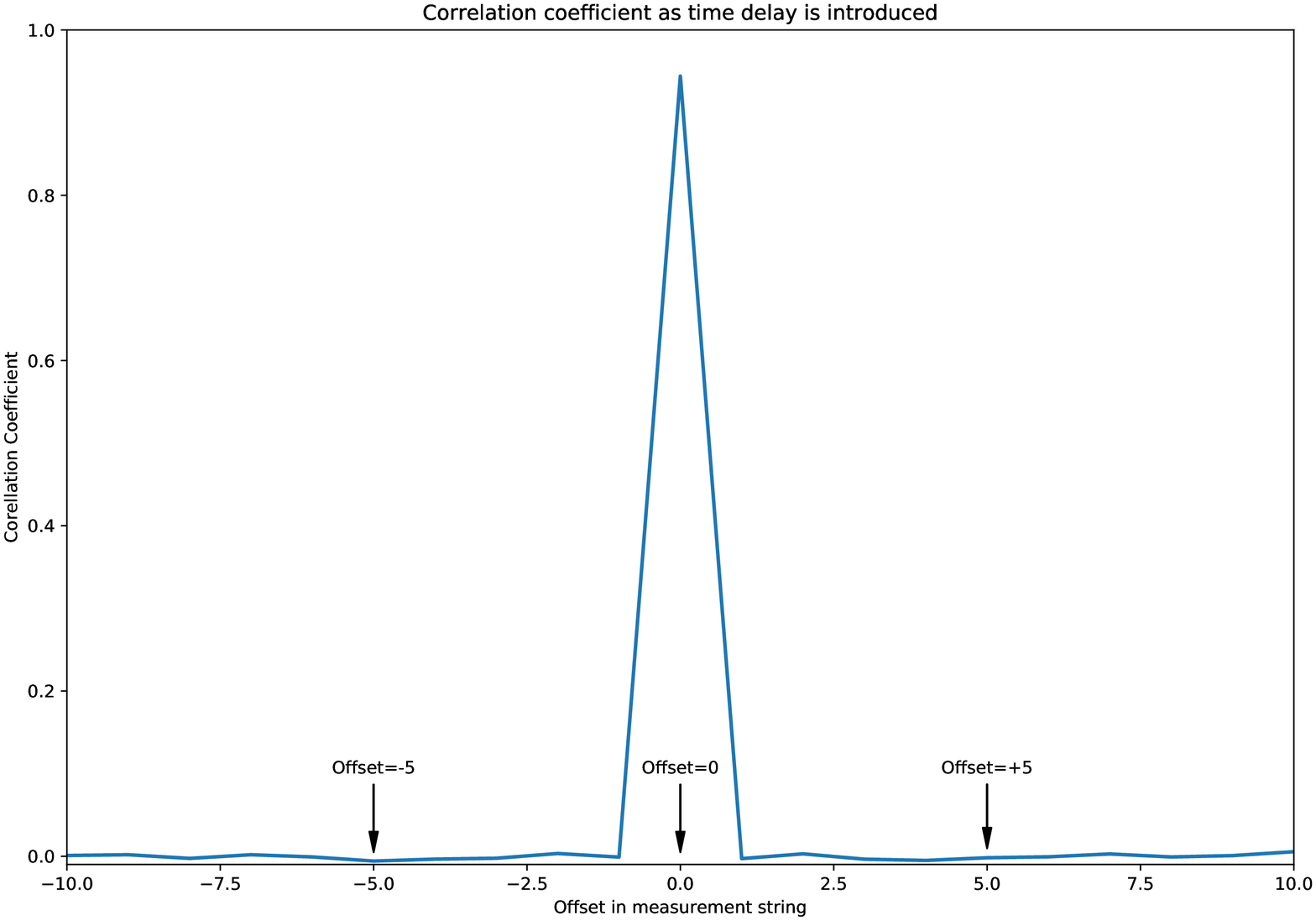}
	\includegraphics[width=1.1\textwidth]{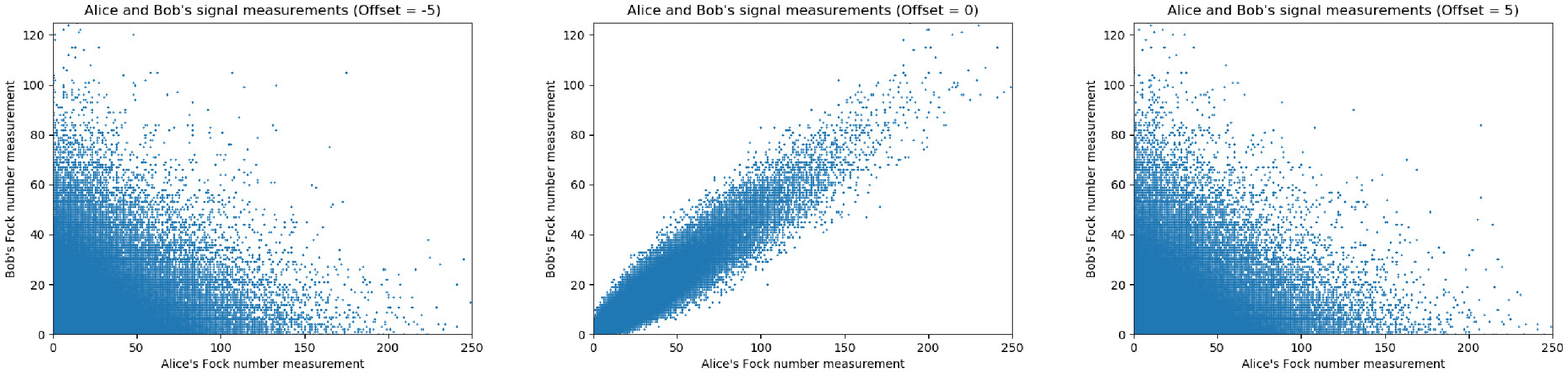}
	\caption{\textbf{Correlation Coefficients.} Changes in the correlation coefficient as a sample set of Alice and Bob's measurement results ($n\approx 100000$) are offset with respect to each other. Correlated measurements are observed in the thermal protocol when time delays are taken into account. Also shown are sample scatter graphs comparing Alice and Bob's measurements, with and without offsetting the data streams. \label{fig:Offset}}
\end{figure}

Using a sample measurement set produced through simulation, we can verify that Alice and Bob are receiving correlated measurements by calculating the correlation coefficient as Alice's data is offset relative to Bob. We can see from Figure \ref{fig:Offset} that the correlations survive beam splitters as expected of thermal sources \cite{HBT}. Offsetting Alice and Bob's data streams shows a clear difference between the correlation coefficients for synchronised measurements and random noise, which would not be observed if a coherent source were used.

Given that we have observed correlations between Alice and Bob's data strings, we now derive bit strings from the Fock state measurements and proceed to calculate Shannon mutual informations to test if a secure key can be produced.

\section{Key Rates} \label{sec:Information-Measurements}

After performing Python simulations, the Shannon mutual information; \(I_{S}\left(A;B\right)\), is calculated using the bit strings produced by each person. This is a measurement of the information gained about one of the involved systems from measurement of the other system.

We begin with the definition for the Shannon entropy for a single system, \(H\left(A\right)=-\sum_{i=0}^{n-1}p_{i}\log_{2}\left(p_{i}\right)\). This describes the uncertainty in predicting the outcome should a measurement be performed on the system where there are n possible measurement outcomes, with outcome $i$ having a probability $p_{i}$ of occurring. For a binary bit string with 0 and 1 being the only possible values, this can be simplified to:
\begin{equation}
H\left(A\right)=-p_{0}\log_{2}\left(p_{0}\right)-\left(1-p_{0}\right)\log_{2}\left(1-p_{0}\right).\label{eq:Binary}
\end{equation}

Here, $p_{0}$ is the probability of measuring the 0 outcome. From this, the mutual information \(I_{S}\left(A;B\right)=H\left(A\right)+H\left(B\right)-H\left(AB\right)\) can be defined, where $H\left(AB\right)$ is calculated by iterating over the four possible outcomes of two people measuring separate bit strings. Once the Shannon entropy is calculated for each bit string, and the mutual information values between the bit strings for each person is measured, we can see if key distribution can be performed. Two classical options for producing usable keys are considered, direct reconciliation and reverse reconciliation.

In direct reconciliation, Alice openly shares additional information in order for corrections to be made to Bob's bit string. For this to produce a secure key, it is required \cite{Bounds} that entropy calculations from the produced bit strings satisfy \(K_{DR}=I_{S}\left(A;B\right)-I_{S}\left(A;E\right)>0\).

Alternatively, reverse reconciliation is the opposite method, where Bob provides the information in order for Alice to make corrections. In this case, successfully creating a secure key requires \cite{Bounds} \(K_{RR}=I_{S}\left(A;B\right)-I_{S}\left(B;E\right)>0\). Therefore, if one of these inequalities are satisfied, a secret key can be produced. The secret key rate K in this case is bounded such that \cite{Key_Rate} \(\max\{K_{DR},\,K_{RR}\}\leq K\left(A;B|E\right)\leq\min\{I_{S}\left(A;B\right),\,I_{S}\left(A;B|E\right)\}\).

Here, \(H(X|Y)=H\left(XY\right)-H\left(Y\right)\) describes the conditional mutual information, the uncertainty in a system, X, given a measurement performed in a second system. So far, the Shannon mutual information values have been used, which can verify security in the case of an individual attack by Eve \cite{25km_QKD}, in which they perform measurements on each pulse sent by Alice before any error correction occurs between Alice and Bob.

If neither of the above reconciliation methods are available, advantage distillation through protocols such as Cascade will allow keys to be produced from the bit strings provided any secrecy is present \cite{Thermal_2}. As measurements have already taken place by this point, this is purely classical error correction. Sharing a random subset of the bit strings allows Alice and Bob to estimate the error rate for use in such algorithms.

While the Shannon entropy calculated through simulation is interesting, it is more useful to analyse the protocol through the von Neumann entropy. Here, we will compare the results of two different methods of calculating von Neumann entropy to sample Shannon entropies produced by the simulation.

\section{Mutual Information and State Variance} \label{sec:Mutual-Information-and}

Performing an analysis similar to that done by Qi et. al. (2017) \cite{Passive} we can calculate von Neumann mutual informations, $I_{N}\left(A;B\right)$. Using Alice's quadrature measurements, Bob's corresponding measurements are estimated, along with Alice's uncertainty on Bob's measurements.
Eve's interception is done with a beam splitter of transmittance $\tau$ and reflectance $\mu$. This gives:

\begin{equation}
\frac{A_{X}}{n_{A}}=\frac{\hat{B}}{\tau n_{B}}
\end{equation}

where $A_{X}$ is one of Alice's measured X quadrature values, and $\hat{B}$ is Alice's estimate of the corresponding quadrature values of the modes at Bob detector. The detector efficiency for Alice, Bob and Eve are given by $n_{A},\:n_{B},$ and $n_{E}$ respectively. By considering asymmetric beam splitters, continuing the analysis shows the quadrature values of the modes received by each person's detectors, $X_{A},$ $X_{B}$, and $X_{E}$ are found:

\begin{equation}
X_{A}=\frac{n_{A}}{2}x_{in}+\sqrt{1-\left(\frac{n_{A}}{2}\right)^2}v_{A}+N_{A},
\end{equation}

\begin{equation}
X_{B}=\frac{\tau n_{B}}{2}x_{in}+\sqrt{1-\left(\frac{\tau n_{B}}{2}\right)^2}v_{B}+N_{B},
\end{equation}

\begin{equation}
X_{E}=\frac{\mu n_{E}}{2}x_{in}+\sqrt{1-\left(\frac{\mu n_{E}}{2}\right)^2}v_{E}+N_{E},
\end{equation}

where $x_{in}$ is the quadrature output from the source, $v_{A},$ $v_{B},$ and $v_{E}$ describe the noise introduced at the beam splitters between the source and each person, and loss at their detector. $N_{A},$ $N_{B},$ and $N_{E}$ describes Gaussian noise added at each person's detector.

Taking the introduced noise to be described by a Gaussian distribution with
mean zero and variance one, we can calculate the uncertainty Alice has on Bob's measurements, then we can perform a similar analysis for Bob and Eve:

\begin{equation}
\Delta_{AB}=\left\langle \left(\hat{B}-X_{B}\right)^{2}\right\rangle =\left(\frac{\tau n_{B}}{n_{A}}\right)^2\left(1-\frac{n_{A}}{2}+\left\langle N_{A}^{2}\right\rangle \right)+1+\left\langle N_{B}^{2}\right\rangle ,
\end{equation}

\begin{equation}
\Delta_{BE}=\left(\frac{\mu n_{E}}{\tau n_{B}}\right)^2\left(1-\frac{\left(\tau n_{B}\right)^2}{2}+\left\langle N_{B}^{2}\right\rangle \right)+1+\left\langle N_{E}^{2}\right\rangle .
\end{equation}

\begin{figure}[H]
	\centering
	\includegraphics[width=0.8\textwidth]{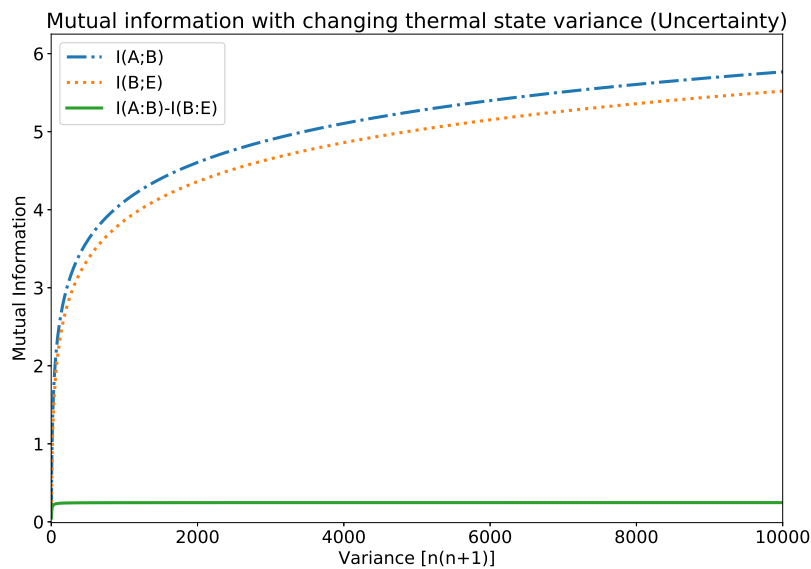}
	\caption{\textbf{Mutual information calculations using uncertainty.} The von Neumann mutual informations $I_{N}\left(A;B\right)$, $I_{N}\left(B;E\right)$ as the variance of the input thermal state, \(\langle n\rangle\left(\langle n\rangle+1\right)\) is changed. This is calculated through uncertainty in Alice and Eve's estimates of Bob's measurements. \label{fig:Mutual Information Calculation}
	}
\end{figure}

The mutual information for Gaussian states can be shown to be \cite{Passive}:
\begin{equation}
I_{N}\left(A:B\right)=\frac{1}{2}\log_{2}\left(\frac{V+\chi}{1+\chi}\right),
\end{equation}

where V is the variance of the input thermal state and $\chi$ is the added noise. If $\chi_{line}$ is the noise added in the channels, and $\chi_{hom}$ is the detection noise, the total added noise in a channel with transmittance T is given by \cite{Passive}:

\begin{equation}
\chi=\chi_{line}+\frac{\chi_{hom}}{T},
\end{equation}

\begin{equation}
\chi_{line}=\frac{1}{T}-2+\Delta,
\end{equation}

\begin{equation}
\chi_{hom}=\frac{1+\left\langle N^{2}\right\rangle }{n_{B}}-1,
\end{equation}

where we have taken $T=1$ and assumed equal detector noise for each person, such that $\left\langle N_{A}^{2}\right\rangle=\left\langle N_{B}^{2}\right\rangle=\left\langle N_{E}^{2}\right\rangle=\left\langle N^{2}\right\rangle=1$. This simplified setup gives $\chi=\Delta$. Figure \ref{fig:Mutual Information Calculation} shows the plots of various mutual information values as variance is adjusted. Eve's beam splitter is assumed to be 50:50.

\section{Covariance Matrix Description} \label{sec:Covariance-Matrices}

To verify this behaviour, we can use a second method to calculate von Neumann entropies, also extending Section \ref{sec:Information-Measurements} to allow for entropy calculations using the quantum state of the system, rather than Shannon entropies of measurements. As the states involved in this protocol are Gaussian, they can be completely described with covariance matrices. For an N-mode state $\rho$, the covariance matrix $\gamma$ is defined as \cite{Raul}:

\begin{equation}
\gamma_{ij}=\Tr\left[\rho\frac{1}{2}\left\{ \left(\hat{r}_{i}-d_{i}\right),\:\left(\hat{r}_{j}-d_{j}\right)\right\} \right].
\end{equation}

Here, \(r=\left(\hat{X}_{1},\hat{P}_{1},...,\hat{X}_{N},\hat{P}_{N}\right)\) consists of a pair of quadrature operators for each mode, $\hat{X}_{i}$ and $\hat{P_{i}}$, with $d_{i}=\left\langle \hat{r}_{i}\right\rangle $ denoting their expectation values. For the inputs into the initial splitter, the covariance matrix $\gamma_{12}$ is given by:

\begin{equation}
\gamma_{12}=\left[\begin{array}{cccc}
V & 0 & 0 & 0\\
0 & V & 0 & 0\\
0 & 0 & 1 & 0\\
0 & 0 & 0 & 1
\end{array}\right]
\end{equation}

Where $V$ is the variance of the quadratures of the beam output by the thermal source. This fully describes the thermal and vacuum modes input into the initial beam splitter. Through applying the beam splitter transformations to the relevant modes, the covariance matrix of the final state is calculated. The transformation, S, for a beam splitter with transmittance $\tau$ and reflectance $\mu$ is given by \cite{Raul}:

\begin{equation}
S\left(\tau ,\mu \right)=\left[\begin{array}{cc}
\tau & \mu \\
-\mu & \tau
\end{array}\right]\otimes I,
\end{equation}

By applying the beam splitters to the appropriate modes, the final covariance matrix can be found:

\begin{equation}
\gamma_{A_{1}A_{2}B_{1}B_{2}E_{1}E_{2}}=\left[\begin{array}{ccc}
\gamma_{A_{1}A_{2}} & C_{AB} & C_{AE}\\
C_{AB}^{T} & \gamma_{B_{1}B_{2}} & C_{BE}\\
C_{AE}^{T} & C_{BE}^{T} & \gamma_{E_{1}E_{2}}
\end{array}\right].
\end{equation}

The sub-matrices are given by:

\begin{equation}
\gamma_{A_{1}A_{2}}=\left[\begin{array}{cc}
\frac{1}{4}\left(V+3\right) & -\frac{1}{4}\left(V-1\right)\\
-\frac{1}{4}\left(V-1\right) & \frac{1}{4}\left(V+3\right)
\end{array}\right]\otimes I,
\end{equation}

\begin{equation}
\gamma_{B_{1}B_{2}}=\left[\begin{array}{cc}
\frac{\tau ^{2}}{4}\left(V+1\right)+\frac{1+\mu ^{2}}{2} & -\frac{\tau ^{2}}{4}\left(V+1\right)+\frac{1-\mu ^{2}}{2}\\
-\frac{\tau ^{2}}{4}\left(V+1\right)+\frac{1-\mu ^{2}}{2} & \frac{\tau ^{2}}{4}\left(V+1\right)+\frac{1+\mu ^{2}}{2}
\end{array}\right]\otimes I,
\end{equation}

\begin{equation}
\gamma_{E_{1}E_{2}}=\left[\begin{array}{cc}
\frac{\mu ^{2}}{4}\left(V+1\right)+\frac{1+\tau ^{2}}{2} & -\frac{\mu ^{2}}{4}\left(V+1\right)+\frac{1-\tau ^{2}}{2}\\
-\frac{\mu ^{2}}{4}\left(V+1\right)+\frac{1-\tau ^{2}}{2} & \frac{\mu ^{2}}{4}\left(V+1\right)+\frac{1+\tau ^{2}}{2}
\end{array}\right]\otimes I,
\end{equation}

\begin{equation}
C_{AB}=\left[\begin{array}{cc}
\frac{\tau}{4}\left(1-V\right) & -\frac{\tau}{4}\left(1-V\right)\\
-\frac{\tau}{4}\left(1-V\right) & \frac{\tau}{4}\left(1-V\right)
\end{array}\right]\otimes I,
\end{equation}

\begin{equation}
C_{AE}=\left[\begin{array}{cc}
-\frac{\mu}{4}\left(1-V\right) & \frac{\mu}{4}\left(1-V\right)\\
\frac{\mu}{4}\left(1-V\right) & -\frac{\mu}{4}\left(1-V\right)
\end{array}\right]\otimes I,
\end{equation}

\begin{equation}
C_{BE}=\left[\begin{array}{cc}
-\frac{\tau \mu}{4}\left(V-1\right) & \frac{\tau \mu}{4}\left(V-1\right)\\
\frac{\tau \mu}{4}\left(V-1\right) & -\frac{\tau \mu}{4}\left(V-1\right)
\end{array}\right]\otimes I.
\end{equation}

Here, $\gamma_{A_{1}A_{2}}$ is the covariance matrix describing the two modes Alice receives at their pair of detectors, with $C_{AB}$ describing covariance between Alice's modes and Bob's. The remaining sub-matrices are similarly defined. From this, von Neumann entropy values are calculated using symplectic eigenvalues. For a covariance matrix $\gamma$, the von Neumann entropy is given by \cite{Raul}:

\begin{equation}
S_{N}\left(\gamma\right)=\sum_{i}G\left(\frac{\lambda_{i}-1}{2}\right)\label{eq:Neumann}
\end{equation}

Where \(G\left(x\right)=\left(x+1\right)\log_{2}\left(x+1\right)-x\log_{2}x\), and $\lambda_{i}$ are the symplectic eigenvalues of $\gamma$. For the covariance matrix for a single mode system, $\gamma_{1}$, this is given by \(\lambda^{2}=\left|\gamma_{1}\right|\). For a two-mode state with the covariance matrix $\gamma_{12}$, taking \(\Delta=\left|\gamma_{1}\right|+\left|\gamma_{2}\right|-2\left|C\right|\) allows the two symplectic eigenvalues, $\lambda_{+}$ and $\lambda_{-}$ to be calculated:

\begin{equation}
\left(\lambda_{\pm}\right)^{2}=\frac{1}{2}\left(\Delta\pm\left[\Delta^{2}-4\left|\gamma_{12}\right|\right]^{\frac{1}{2}}\right).
\end{equation}

\begin{figure}[H]
	\centering
	\includegraphics[width=0.8\textwidth]{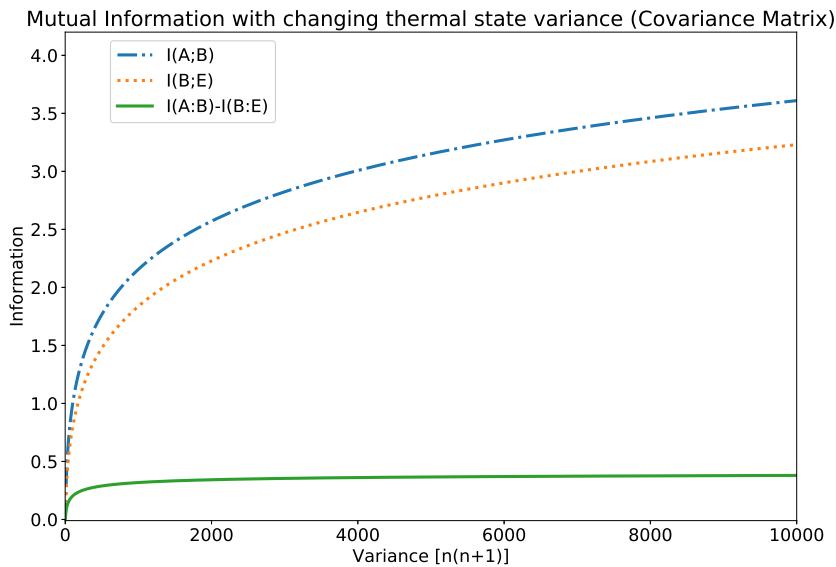}
	\caption{\textbf{Mutual information calculations using covariance.} Von Neumann mutual information calculations plotted against thermal state variance,
		found using the covariance matrix of the final state.
		Here, Eve performs interception using a 50:50 beam splitter. \label{fig:Covariance Graph}
	}
\end{figure}

Mutual informations calculated in this way can be plotted against variance, this is displayed in Figure \ref{fig:Covariance Graph}. Upper and lower bounds can be placed on the mutual information values in the same manner as when Shannon entropies were used. However, requiring \(K_{RR}=I_{N}\left(A;B\right)-I_{N}\left(B;E\right)>0\), where $I_{N}\left(A;B\right)=S_{N}\left(\gamma_{A}\right)+S_{N}\left(\gamma_{B}\right)-S_{N}\left(\gamma_{AB}\right)$, allows for security against a stronger set of attacks. In the case of these "collective attacks", Eve does not perform measurements until after classical communication between Alice and Bob has occurred. In the example shown in Figure \ref{fig:Covariance Graph}, it can be seen that as the variance of the thermal state is increased, the protocol remains secure in the case Eve uses a 50:50 beam splitter. In this case, reverse reconciliation is used as $K_{RR}$ is positive. Additionally, analysing either the covariance matrix or measurement uncertainty both produce mutual information graphs which follow similar patterns as variance is increased.

With two methods of calculating von Neumann entropy displaying similar behaviour, we may now compare the outputs to Shannon entropies calculated through the simulation.

\section{Results} \label{sec:Results}

\begin{figure}[H]
	\centering
	\includegraphics[width=0.8\textwidth]{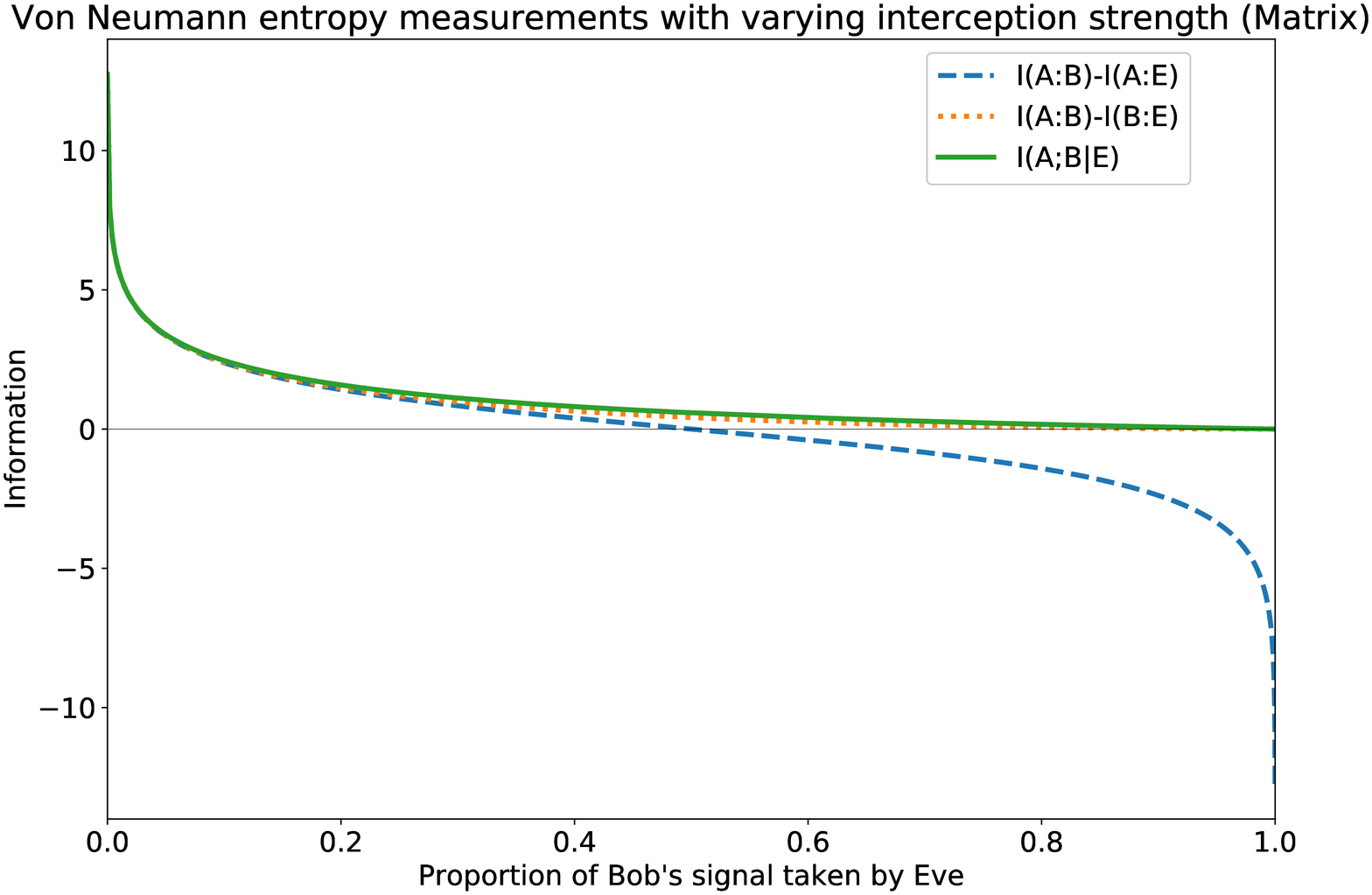}
	\includegraphics[width=0.8\textwidth]{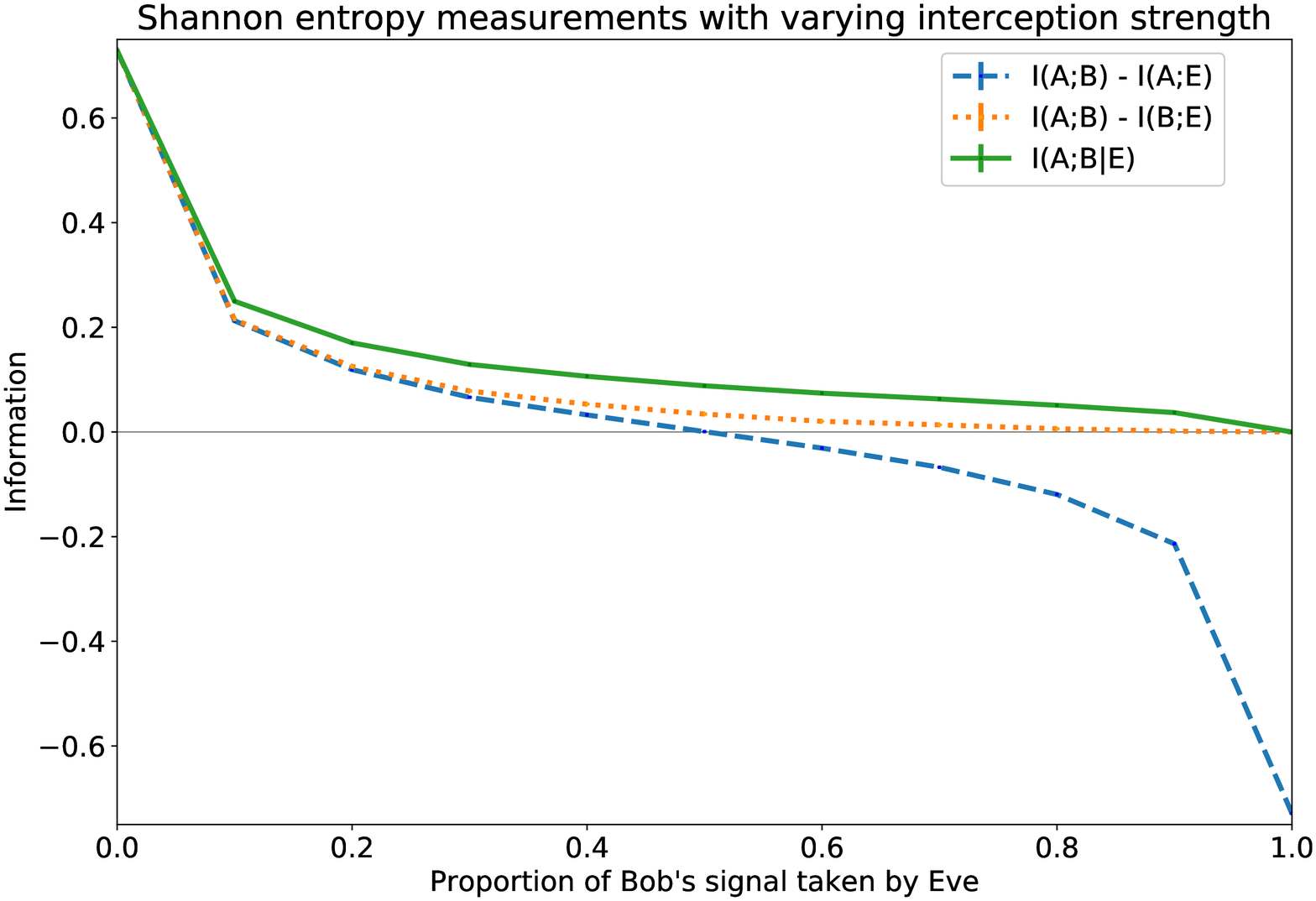}
	\caption{\textbf{Information with varying interception strength.} Calculations of von Neumann entropy using the covariance matrix and measurements of Shannon entropy taken from the simulation as the transmittance of Eve's beam splitter is varied. The notable result is that $I\left(A;B\right)-I\left(B;E\right)>0$ holds in both cases provided Bob receives a nonzero proportion of the signal sent to him. This used an average photon number of 200. Error bars describe one standard deviation. Due to discord in the system, von Neumann entropies have larger magnitudes than Shannon entropies. \label{fig:Measurements}}
\end{figure}

We performed calculations of the Shannon entropy for the three bit strings and the mutual information between each pair of strings. These strings were produced through the Python simulation. Currently no loss or noise is considered, this allows the simulation to be performed with a lossless Eve. The transmittance of Eve's splitter is varied to measure the effect of Eve's interception strength on the mutual information values. Also calculated was the von Neumann entropy of each mode and pair of modes, using equation \ref{eq:Neumann} and the covariance matrix describing the final state of the system. The results of these calculations and measurements are shown in Figure \ref{fig:Measurements}.

Meeting the restrictions placed on von Neumann mutual information which ensure secrecy in the case of a collective attack, $I_{N}\left(A;B\right)-I_{N}\left(B;E\right)>0$ or $I_{N}\left(A;B\right)-I_{N}\left(A;E\right)>0$, allow the protocol to be secure against a greater range of attacks than the restrictions based on Shannon entropy. Both are included here so that it can be seen that changes in von Neumann entropies are reflected in the Shannon entropies of the bit strings derived by each person after the protocol has been carried out.

It can be seen from Figure \ref{fig:Measurements} that $I\left(A;B\right)-I\left(A;E\right)$ crosses zero in both cases when Eve's beam splitter reflects half of the beam sent to Bob. This is expected as Bob and Eve's positions in the protocol are interchangeable in this special case, so $I\left(A;B\right)=I\left(A;E\right)$. If over half of Bob's beam is reflected by Eve, a key cannot be produced via direct reconciliation. However, the second possible requirement of $I\left(A;B\right)-I\left(B;E\right)>0$ is always satisfied in the no-loss scenario provided Eve's interception beam splitter has nonzero transmittance. This means that reverse reconciliation may be used to produce a secret key and establish secure communication during collective or individual attacks. This would allow the thermal state central broadcast protocol to be used as a method of quantum key distribution.
It is also clear that the von Neumann entropy has a higher magnitude than Shannon, this is expected due to the presence of discord in the system, which the Shannon entropy does not consider.

\section{Conclusions}

When considering a system without loss and noise, the lower bound placed on the key rate when reverse reconciliation is considered is positive under a beam splitter attack even when Eve has zero loss. This would allow for a secret key to be produced between Alice and Bob, and therefore secure communication could take place in the presence of an eavesdropper performing collective or individual attacks. Additionally, two separate methods of von Neumann mutual information analysis both showed that thermal sources with higher variance than those that could be produced in the Monte Carlo simulation allowed for a key to be produced with little change in the lower bound of the key rate.

Future work in this area could focus on examining the effects of adding noise and loss into different channels, checking if a positive key rate could be maintained. This is especially relevant for thermal states due to added noise being a large barrier to successful QKD. Additionally, a practical setup following the diagram shown in Figure \ref{fig:Protocol} would allow the protocol to be performed experimentally. This allows real key rates to be measured and would show if the protocol continues to be functional when using thermal sources likely to be employed in modern communication.

\section*{Acknowledgements}

Work was undertaken on ARC4, part of the High Performance Computing facilities at the University of Leeds, UK. DJ is supported by the Royal Society and also a University Academic Fellowship. This work was supported by the Northern Triangle Initiative Connecting capability fund as well as funding from the UK Quantum Technology Hub for Quantum Communications Technologies EP/M013472. Data used to plot the Shannon entropy graph in Figure \ref{fig:Measurements} is available from the Research Data Leeds Repository with the identifier https://doi.org/10.5518/944 \cite{doinumber}.

\section*{References}

\bibliographystyle{unsrt}
\bibliography{References}

\end{document}